# Impact of the Out-of-Plane Flow Shear on Magnetic Reconnection at the Flanks of Earth's Magnetopause

Haoming Liang[1,2], Li-Jen Chen[2], Naoki Bessho[1,2], Jonathan Ng[1,2]

1. University of Maryland College Park, 2. NASA-GSFC

**Abstract**

Magnetic reconnection changes the magnetic field topology and facilitates the energy and particle exchange at magnetospheric boundaries such as the Earth's magnetopause. The flow shear perpendicular to the reconnecting plane prevails at the flank magnetopause under southward interplanetary magnetic field (IMF) conditions. However, the effect of the out-of-plane flow shear on asymmetric reconnection is an open question. In this study, we utilize kinetic simulations to investigate the impact of the out-of-plane flow shear on asymmetric reconnection. By systematically varying the flow shear strength, we analyze the flow shear effects on the reconnection rate, the diffusion region structure, and the energy conversion rate. We find that the reconnection rate increases with the upstream out-of-plane flow shear, and for the same upstream conditions, it is higher at the dusk side than at the dawn side. The diffusion region is squeezed in the outflow direction due to magnetic pressure which is proportional to the square of the Alfvén Mach number of the shear flow. The out-of-plane flow shear increases the energy conversion rate $\mathbf{J} \cdot \mathbf{E}'$, and for the same upstream conditions, the magnitude of $\mathbf{J} \cdot \mathbf{E}'$ is larger at the dusk side than at the dawn side. This study reveals that out-of-plane flow shear not only enhances the reconnection rate but also significantly boosts energy conversion, with more pronounced effects on the dusk-side flank than on the dawn-side flank. These insights pave the way for better understanding the solar wind-magnetosphere interactions.

**Key Points**
  (1) The reconnection rate increases with the out-of-plane flow shear and is higher at dusk than at dawn side for the same upstream conditions
  (2) The diffusion region is squeezed in the outflow direction, causing its length to decrease as the out-of-plane shear flow increases
  (3) The energy conversion rate increases with flow shear strength and is higher at dusk than at dawn side for the same upstream conditions

## 1. Introduction

Magnetic reconnection is a crucial process driving the exchange of energy and particles at the boundary of Earth's magnetosphere. At the flanks of Earth's magnetopause, reconnection events are particularly interesting due to their ability to transfer energy, mass, and momentum from the solar wind into the magnetosphere.

Under the southward interplanetary magnetic field (IMF) condition, the potential areas for the locations where magnetic reconnection occurs can be significantly extended across the magnetopause. (e.g., Scurry et al., 1994; Phan et al., 1996; Phan et al., 2000; Fuselier et al., 2005; Trattner et al., 2007; Hasegawa et al., 2016). The reconnection configuration at the flanks of Earth's magnetopause under the southward IMF condition is shown in Figure 1. The plasma bulk flow at the magnetosheath (msh) side normal to the reconnection plane, i.e., the *Y-Z* plane in Geocentric Solar Ecliptic (GSE) coordinates, can be significant relative to the flow on the magnetosphere (msp) side. In the frame of the magnetospheric flow, the magnetosheath flow $u_{msh}$ determines the out-of-plane flow shear in the reconnection sites. There are two reconnection scenarios: (a) at the dusk-side flank, the current direction is the same as $u_{msh}$; (b) at the dawn-side flank, the current direction is opposite to $u_{msh}$. We are interested in investigating the two scenarios because it is unclear (1) whether and how the upstream out-of-plane flow affects asymmetric reconnection, and (2) whether the relative directions between the upstream magnetosheath flow and the current carrying flow can produce different effects to the reconnection at the dusk and the dawn sides. Asymmetric reconnection with the out-of-plane flow at the flanks has been observed by spacecraft during magnetopause crossings (e.g., Souza et al., 2017). It is also observed either during the dayside X-line spreading (e.g., Zou et al., 2018) or within the Kelvin-Helmholtz Instability (KHI) vortices (Gurram et al., 2024).

The effects of the out-of-plane flow shear on reconnection under various upstream conditions have been explored using numerical simulations. By using two-dimensional (2D) resistive MHD, Wang et al. (2008) found that the flow shear can generate quadrupolar out-of-plane magnetic perturbation without the Hall effect in symmetric reconnection. Wang et al. (2012) studied the distorted quadrupolar out-of-plane magnetic field due to the flow shear and the Hall effect by using 2D Hall MHD. Chen et al. (2013) studied the magnetic flux rope generation due to the flow shear by using three-dimensional (3D) Hall MHD. Wang et al. (2015) studied the effect of symmetric and antisymmetric out-of-plane flow shears on asymmetric reconnection by using 2D hybrid simulation. They found that the flow shear can increase the reconnection rate, distort the quadrupolar Hall field patterns, and generate secondary islands. Liu et al. (2018) used 2D and 3D particle-in-cell (PIC) simulations to study the super-Alfvénic flow shear effect on positron-electron symmetric reconnection and found that the reconnection rate is ~0.1 while a reversed current at the X-line is observed which is impossible in resistive MHD. Nakamura et al. (2020) studied reconnection at the flank during KHI. They found that under the southward IMF condition, reconnection can occur in the plane perpendicular to the KHI plane and the reconnection leads to a quick decay of the vortex structures. They used their results to interpret the difference of the observation occurrence of the magnetopause KH vortices between northward and southward IMFs.

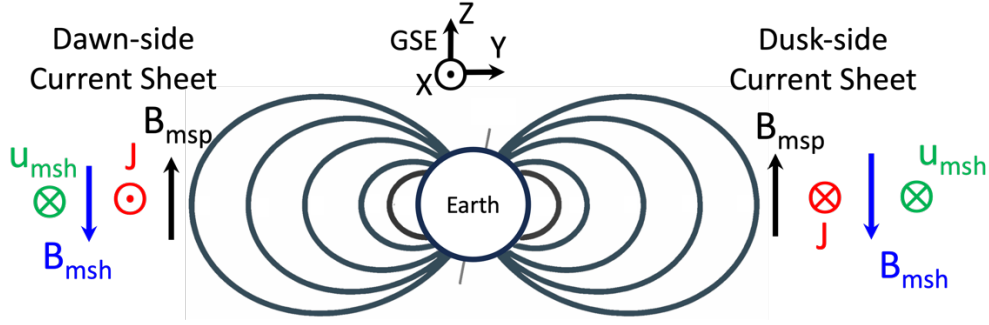

*Figure 1 Diagram of the reconnection sites with respect to the Earth's dipole fields during southward IMF. The view is from the sun. The directions of magnetosphere (magnetosheath) magnetic field $B_{msp}(B_{msh})$, current density J, and magnetosheath flow $u_{msh}$ are shown for both dawn-side and dusk-side current sheets.*

In this work, we study the impact of the out-of-plane flow shear on asymmetric reconnection at both dusk-side and dawn-side flanks. By using PIC simulations, we investigate the reconnection rate, the diffusion region configuration, and the energy conversion rate under different strengths of the out-of-plane magnetosheath flow: no flow, sub-Alfvénic flow, and super-Alfvénic flow. This manuscript is structured as follows: Section 2 outlines the methodology, including the simulation parameters and setup. Section 3 present the results, offering a detailed analysis of the reconnection rates, the diffusion region configuration, and the energy conversion rate. Finally, we discuss our findings and propose directions for future research.

## 2. Simulation Setup

To investigate the impact of the out-of-plane flow shear on asymmetric reconnection, we carry out 2.5D PIC simulations using the p3d code (Zeiler et al., 2002). Here, 2.5D means 2D in position space and 3D for fields and velocity. The code uses the relativistic Boris particle stepper (Birdsall and Langdon, 2018) for the particles and the trapezoidal leapfrog method (Guzdar et al., 1993) on the electromagnetic fields. The divergence of the electric field is cleaned every ten particle time steps using the multigrid approach (Trottenberg and Clees, 2009). The normalization is based on an arbitrary magnetic field strength $B_0$ and density $n_0$. Spatial and temporal scales are normalized to the ion inertial length $d_i = c/\omega_{pi}$ and the ion cyclotron time $\Omega_{ci}^{-1}$, respectively, where $\omega_{pi} = \left(\frac{4\pi n_0 e^2}{m_i}\right)^{1/2}$ is the ion plasma frequency and $\Omega_{ci} = eB_0/m_i c$ is the ion cyclotron frequency, the unit charge $e$ and the ion mass $m_i$ are set as 1 in the simulation. Thus, velocities are normalized to the Alfvén velocity $v_{A0} = d_i \Omega_{ci}$. Electric fields are normalized to $v_{A0}B_0/c$. Note that the equations in the rest of this section, originally in cgs units, have been converted to code units (i.e., using the normalized variables). For example, the Gauss' law in cgs, $4\pi\rho_e^{(G)} = \nabla^{(G)} \cdot \boldsymbol{E}^{(G)}$, where "(G)" denotes the variables in cgs, $\rho_e$ is charge density and $\boldsymbol{E}$ is electric field, is converted to the form with the normalized variables, $4\pi e n_0 \rho_e = \frac{v_{A0}B_0}{c d_i} \nabla \cdot \boldsymbol{E} \Rightarrow \rho_e = \frac{v_{A0}^2}{c^2} \nabla \cdot \boldsymbol{E}$, where $\rho_e^{(G)} = en_0\rho_e$, $\boldsymbol{E}^{(G)} = \frac{v_{A0}B_0}{c}\boldsymbol{E}$, and $\nabla^{(G)} = \frac{1}{d_i}\nabla$.

The simulation domain is $l_x \times l_y = 51.2 \times 25.6$ with periodic boundary conditions in every direction. Note that the simulation coordinates are different from the GSE coordinates in Figure 1. A double current sheet initial condition is used, with a magnetic field given by

$$B_x(y) = \frac{1}{2}(b_1 + b_2)\left(\tanh\frac{y - \frac{1}{4}l_y}{w_0} - \tanh\frac{y - \frac{3}{4}l_y}{w_0} - 1\right) + \frac{1}{2}(b_2 - b_1),$$

where the subscript "1" ("2") corresponds to the magnetosphere (magnetosheath) side, $b_{1,2}$ is the magnetic field magnitude, and $w_0 = 0.5$ is the initial half-thickness of the current sheet. We use $b_1 = 1.5$ and $b_2 = 0.5$. The out-of-plane flow is given by

$$u_z(y) = \frac{1}{2}(u_1 + u_2)\left(\tanh\frac{y - \frac{1}{4}l_y}{w_0} - \tanh\frac{y - \frac{3}{4}l_y}{w_0} - 1\right) + \frac{1}{2}(u_2 - u_1),$$

where the magnetosphere flow $u_1 = 0$ and the magnetosheath flow $u_2 = u_{msh}$, because we only consider the magnetosheath flow for simplification. The initial advection electric field is calculated as $E_y(y) = -u_z(y)B_x(y)/c$. Due to non-zero $\frac{\partial E_y}{\partial y}$ across the current sheet, according to Gauss' law, $\rho_e = \frac{v_{A0}^2}{c^2}\frac{\partial E_y}{\partial y}$, and a finite charge density $\rho_e = \delta n_i - \delta n_e$ disrupts the quasi-neutrality. Similar to Liu et al. (2018), we assume $\delta n_i = -\delta n_e$; as a result, $\delta n_i = -\delta n_e = \frac{v_{A0}^2}{2c^2}\frac{\partial E_y}{\partial y}$. The number density can be calculated as

$$n_{i,e}(y) = \frac{1}{2}(n_2 - n_1)\left(\tanh\frac{y - \frac{1}{4}l_y}{w_0} - \tanh\frac{y - \frac{3}{4}l_y}{w_0} - 1\right) + \frac{1}{2}(n_1 + n_2) + \delta n_{i,e}$$

Here, $c/v_{A0} = 15$ is the speed of light in the simulation. We use $n_1 = 2/3$ and $n_2 = 4/3$. In the case of the highest-speed flow, $\delta n_{i,e}$ is no more than 2% of the local density, which is less than the intrinsic fluctuations of density (~3.5%) in PIC simulation. This disruption of quasi-neutrality thus disappears quickly when the simulation starts. Note that in reality, the finite charge density could be a factor of 400 smaller than our simulation if we consider that the speed of light is about 300 times the Alfvén speed at the magnetopause. Assuming the ion-to-electron temperature ratio $T_i(y)/T_e(y) = 2$ everywhere and considering the asymptotic temperature $T_{e1} = T_{e2} = 0.5$, the temperature profiles can be determined by the pressure balance

$$n_i(y)T_i(y) + n_e(y)T_e(y) + \frac{B_x^2(y)}{2} - \frac{v_{A0}^2 E_y^2(y)}{2c^2} = \text{const},$$

where the first two terms are ion and electron pressures, and the third and fourth terms are magnetic and electric pressures (e.g., Liu et al., 2018). The initial velocity distribution functions are drifting Maxwellians for both electrons and ions. The ion-to-electron mass ratio $m_i/m_e = 25$. Initially, in each grid cell, there are 400 macro electrons and 400 macro ions, each weighted according to the local density. The smallest spatial scale is the Debye scale on the magnetosheath side, $\lambda_{De,msh} = 0.041$ and we choose the spatial grid size $dx = dy = 0.025$. Two reconnection sites are perturbed at the upper and lower current sheets as shown in Figure 2. The upper (lower) current sheet mimics the dusk-side (dawn-side) reconnection since the current is parallel (anti-parallel) to the magnetosheath flow.

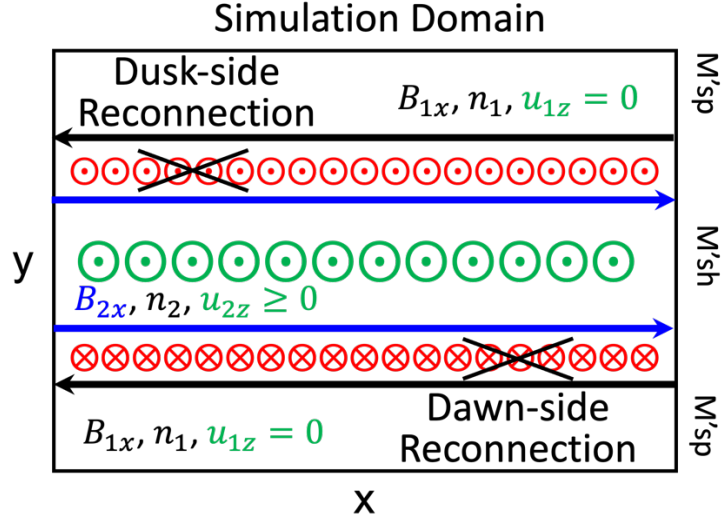

*Figure 2 Simulation domain with the directions of the initial magnetic field (black for magnetosphere "M'sp" and blue for magnetosheath "M'sh"), out-of-plane flow (green), and current density (red). The locations of dawn-side and dusk-side reconnection are shown as the crossing lines.*

In this study, we vary the magnetosheath flow magnitudes without changing other parameters. In our three simulation runs, $u_{msh} = 0, 0.3$, and $1$, respectively. Normalized to the magnetosheath Alfvén speed ($v_{A2} \approx 0.43$), they correspond to Alfvén Mach numbers $M_A = 0, 0.7$, and $2.3$, respectively. They correspond to no flow, sub-Alfvénic flow, and super-Alfvénic flow in the magnetosheath, respectively. These Alfvén Mach numbers are used to name the three simulation runs throughout this paper. For example, we name the dusk-side reconnection under $M_A = 2.3$ magnetosheath flow as the $M_A = 2.3$ (dusk) case. For the $M_A = 0$ run, the upstream conditions are the same for the upper and lower CSs, and in order to avoid confusion, we thus only analyze the dusk-side case, named as $M_A = 0$ case. Note that if we normalized the out-of-plane magnetosheath flow to the predicted outflow Alfvén speed ($\approx 0.80$) according to Cassak and Shay (2007), the Alfvén Mach number $M_A^{CS07} = 0, 0.375$, and $1.25$, respectively.

It is worth noting that the super-Alfvénic magnetosheath flow case does not represent a super-Alfvénic flow shear (e.g., Liu et al., 2018). Ma et al. (2016) pointed out that in a certain moving frame, when the shear flows on both sides of the inflow region are greater than the inflow fast mode speed $\sqrt{v_A^2 + c_s^2}$, then the reconnection layer will form an expanding outflow region to maintain the total pressure balance, where $v_A$ and $c_s$ are Alfvén speed and sound speed, respectively. In our study, the fast mode speeds on the magnetosheath and magnetosphere sides are 1.20 and 1.68, respectively, which means that our highest flow shear case still corresponds to the sub-fast/sub-Alfvénic shear.

## 3. Simulation Results

We present the simulation results about the reconnection rate, the ion diffusion region configuration, the upstream plasma parameters, and the energy conversion.

## 3.1 Reconnection Rate

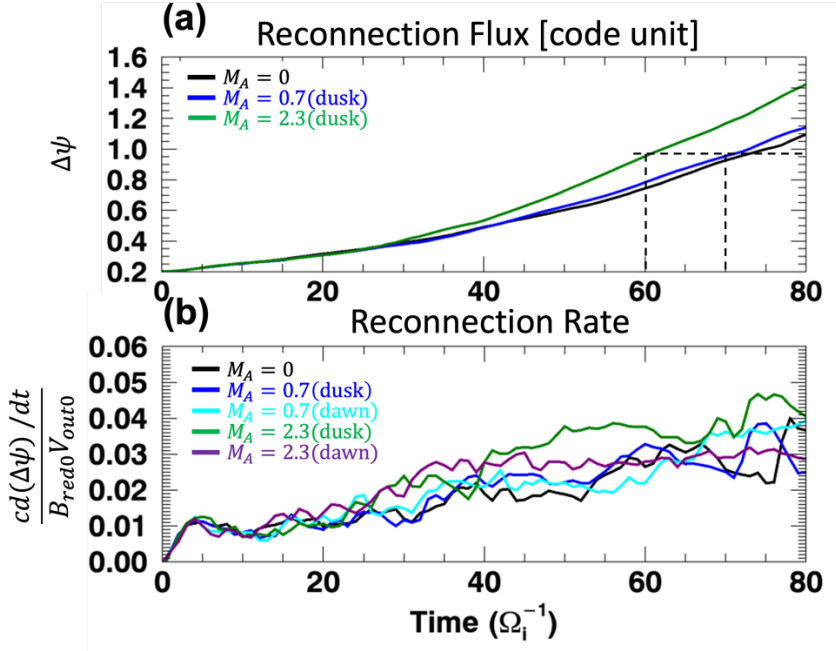

Figure 3 (a) Reconnection flux and (b) reconnection rate for different cases. The reconnection flux is in code unit. The reconnection rate is normalized to the reduced magnetic field $B_{red0}$ and outflow Alfvén speed $v_{out0}$ (Cassak and Shay, 2007) based on asymptotic magnetic fields and densities. The vertical dashed lines in (a) indicate the times selected for the analysis in this paper, where the $M_A = 2.3$ (dusk) case at t=60 has the same reconnected flux as the $M_A = 0.7$ (dusk) and $M_A = 0$ cases at t=70.

The reconnection flux and rate are shown in Figure 3. The reconnection flux $\Delta\psi$ is calculated as the difference between the magnetic flux at the X-point and the O-point. The reconnection rate is normalized to the reduced magnetic field $B_{red0} = \frac{2b_1 b_2}{b_1 + b_2}$ and the predicted outflow velocity $v_{out0} = \left[\frac{b_1 b_2 (b_1 + b_2)}{m_i(n_1 b_2 + n_2 b_1)}\right]^{1/2}$ (Cassak and Shay, 2007). The subscript "0" indicates the usage of the asymptotic values instead of the values upstream of the diffusion region. During a relatively steady state within t=50-70, the $M_A = 2.3$ (dusk) case has the highest reconnection rate among all the runs. Note that the $M_A = 0.7$ (dawn) run only shows steady signatures in the interval between 50-60. We thus analyze the steady state of the $M_A = 0.7$ (dawn) case at t=60. For the $M_A = 0$ run, we do not distinguish dawn or dusk reconnection, because the upstream conditions are the same in this run.

The averaged reconnection rates over t=50-70 are 0.026 for both the $M_A = 0$ case and $M_A = 0.7$ (dawn) case (averaged over t=50-60), 0.027 for the $M_A = 0.7$ (dusk) case, 0.028 for the $M_A = 2.3$ (dawn) case, and 0.036 for the $M_A = 2.3$ (dusk) case. The key questions to answer in this paper are (1) why does the $M_A = 2.3$ (dusk) have a higher reconnection rate than the $M_A = 0$ and $M_A = 0.7$ (dusk) cases? (2) why does the $M_A = 2.3$ (dawn) have a lower rate than $M_A = 2.3$ (dusk)? In order to answer the questions, we further analyze the results at specific times. In the following sections, for the $M_A = 2.3$ run, we will analyze the results at t=60; for the $M_A = 0$ and $M_A = 0.7$ (dusk) cases, we will analyze the results at t=70 when the reconnection flux is the same as that at t=60 for the $M_A = 2.3$ (dusk), as indicated by the dashed lines shown in Figure 3(a).

3.2 Ion Diffusion Region (IDR)

To understand the reconnection rates, we investigate the aspect ratio $\delta/L$ of the IDR, where $\delta$ ($L$) is the half-width of IDR along the inflow (outflow) direction. We use the total slippage of the ion species $|\Delta \boldsymbol{u}_{\perp i}| = |\boldsymbol{u}_i - (\boldsymbol{u}_i \cdot \boldsymbol{B})\boldsymbol{B}/B^2 - c(\boldsymbol{E} \times \boldsymbol{B})/B^2|$ (Schindler et al., 1991; Goldman et al., 2016) to identify IDR, where $\boldsymbol{u}_i$ is ion bulk flow velocity, and $\boldsymbol{E}$ and $\boldsymbol{B}$ are the electric and the magnetic field, respectively. The spatial distributions of $|\Delta \boldsymbol{u}_{\perp i}|$ for all cases are shown in Figure 4. The position $(x_0, y_0)$ indicates the X-point location at the time selected for analysis for each case, which has been discussed in Section 3.1. Note that the X-point is slowly drifting during the simulations, while the peak value of the $|\Delta \boldsymbol{u}_{\perp i}|$ is not sensitive to the drifting X-point location because it is determined by ion kinetic scale physics. Therefore, in Figure 4, the $x - x_0 = 0$

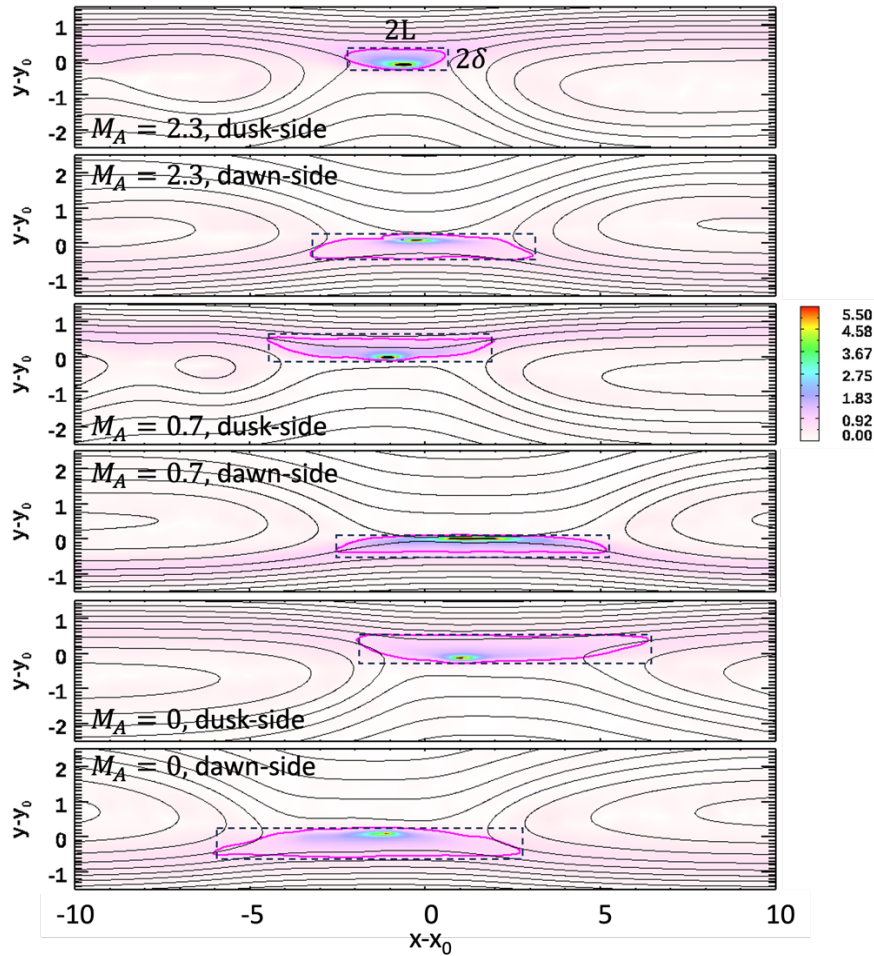

Figure 4 2D plots of total slippage of the ion species $|\Delta \boldsymbol{u}_{\perp i}|$. The magenta line is the contour at 10% of $|\Delta \boldsymbol{u}_{\perp i}|$ peak value and it denotes an approximate ion diffusion region. The dashed boxes are used to estimate the aspect ratios. $x_0$ and $y_0$ are the coordinates of the X-point at the time selected for analysis for each case.

location is not aligned with the peak value of the $|\Delta \boldsymbol{u}_{\perp i}|$. The magenta line is the contour at 10% of the $|\Delta \boldsymbol{u}_{\perp i}|$ peak value, which denotes an approximate IDR. The dashed boxes are used to estimate the aspect ratios, which show $\delta/L(M_A = 2.3, \text{dusk}) \approx 1/5$, $\delta/L(M_A = 2.3, \text{dawn}) \approx \delta/L(M_A = 0.7, \text{dusk}) \approx 1/8$, and $\delta/L(M_A = 0.7, \text{dawn}) \approx \delta/L(M_A = 0, \text{dusk}) \approx \delta/L(M_A = 0, \text{dawn}) \approx 1/10$. The aspect ratios are approximately proportional to the reconnection rates for all these cases discussed in Section 3.1.

The high aspect ratio for the $M_A = 2.3$ (dusk) case is mainly due to a short $L$. Similar to the interpretation in Liu et al. (2018), the short $L$ is because of a squeeze of the magnetic pressure in the outflow region. Figure 5a, similar to Figure 2 in Liu et al. (2018) for symmetric reconnection, shows how the out-of-plane magnetic field component is increased due to the out-of-plane flow dragging. The scenario aligns with the findings of La Belle-Hamer et al. (1995), who demonstrated

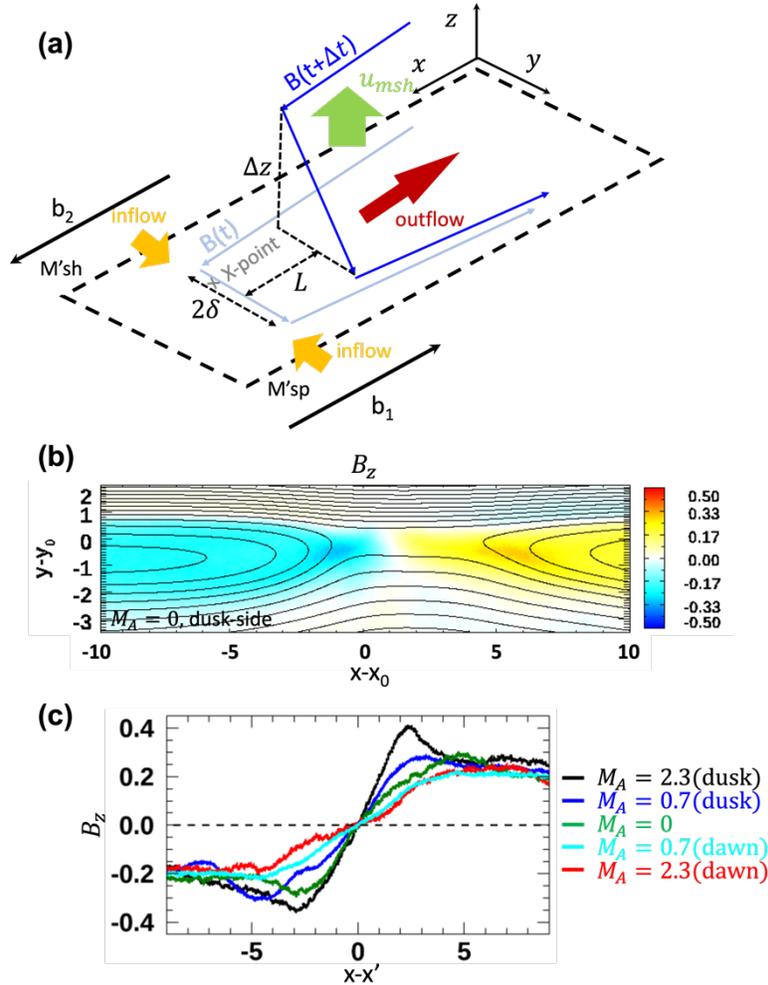

Figure 5 (a) Diagram describing the motion of reconnected magnetic field line due to the out-of-plane flow $u_{msh}$ from Time $t$ to Time $t + \Delta t$. Similar to Figure 2 in Liu et al (2018). (b) 2D plot of the out-of-plane component $B_z$ in the $M_A = 0$ (dusk) case. (c) The profiles of $B_z$ at a cut through y=-0.8 for all cases. The $x'$ is the location of $B_z = 0$ for each case.

that reconnected magnetic field lines are dragged into opposite directions on the two sides of the outflow region, forming a perpendicular magnetic field component ($B_z$ in this study) due to the frozen-in condition. The outflow speed is approximately the outflow Alfvén speed $v_{out0}$. A field line propagates from the X-point to the edge of the diffusion region, which takes time $\Delta t \sim \frac{L}{v_{out0}}$.

In the meantime, the field line is dragged toward the out-of-plane direction with a speed $u_{msh}$, and the distance of the drag is $\Delta z \sim u_{msh} \Delta t \sim u_{msh} \frac{L}{v_{out0}}$. Because of the drag, the generated out-of-plane component $B_{z,out}$ has a relation $\frac{B_{y,out}}{B_{z,out}} \sim \frac{2\delta}{\Delta z} \sim \frac{\delta_{X,msh}+\delta_{X,msp}}{\Delta z}$, where the subscript "out" denotes the outflow, $\delta_{X,msh}$ ($\delta_{X,msp}$) is the distance from the X-point to the upstream magnetosheath (magnetosphere) (Cassak and Shay, 2007). Due to the field line geometry around the X-point, we have $\frac{b_1}{L} \sim \frac{B_{y,out}}{\delta_{X,msp}}$ and $\frac{b_2}{L} \sim \frac{B_{y,out}}{\delta_{X,msh}}$, which are similar to Liu et al. (2018) for symmetric reconnection. Thus, $B_{z,out} \sim \frac{B_{y,out}\Delta z}{\delta_{X,msh}+\delta_{X,msp}} \sim \frac{B_{y,out} u_{msh} \frac{L}{v_{out0}}}{L\frac{B_{y,out}}{b_2}+L\frac{B_{y,out}}{b_1}} \sim \frac{u_{msh}}{v_{out0}} \frac{b_1 b_2}{b_1+b_2}$, which provides significant magnetic pressure ($\propto \left(\frac{u_{msh}}{v_{out0}}\right)^2$) to squeeze the IDR length $L$ when $u_{msh}$ is large. As discussed in Section 2, in this study, $v_{out0} \approx 0.80$; for the $M_A = 0, 0.7,$ and 2.3 cases, $\frac{u_{msh}}{v_{out0}} = M_A^{CS07} = 0, 0.375,$ and 1.25, and $\left(\frac{u_{msh}}{v_{out0}}\right)^2 = 0, 0.14, 1.56$, respectively. This explains why the aspect ratios and reconnection rates for $M_A = 0$ and $M_A = 0.7$ are close to each other, but significantly different than those for the $M_A = 2.3$ (dusk) case. Figure 5(b) shows $B_z$ in the $M_A = 0$ (dusk) case. Without the out-of-plane flow, the $B_z$ in the $M_A = 0$ case represents the Hall magnetic field (Pritchett, 2008). The peak values of the $B_z$ are located in the outflow regions on the magnetosheath side and exhibit opposite signs. The out-of-plane flow dragging, as shown in Figure 5(a), is a different mechanism for generating the $B_z$ component compared to the mechanism that generates the Hall magnetic field. Figure 5(c) shows the $B_z$ component at a cut along the $x$ axis through $y = -0.8$ (i.e., slightly below the X-point on the magnetosheath side) for all cases. The profiles are aligned with the location $x'$ where the $B_z = 0$ for each case. For the $M_A = 0$ case, $B_z$ is the Hall magnetic field. For the other cases, $B_z$ includes both the Hall field and the dragged field components. For the dusk-side (dawn-side) cases, the dragged field is in the same (opposite) direction as the Hall field, which increases (decreases) $B_z$ and the increment (reduction) is proportional to the out-of-plane flow speed at $|x - x_0| < 5$. These simulation results thus validate our hypothesis in Figure 5.

## 3.3 Upstream conditions

In order to understand the different reconnection between $M_A = 2.3$ (dawn) and $M_A = 2.3$ (dusk), we investigate the upstream conditions for the $M_A = 0$, $M_A = 0.7$ (dusk), $M_A = 2.3$ (dusk), and $M_A = 2.3$ (dawn) cases. Figure 6 shows the reconnecting magnetic field $B_x$, density $n_i$, ion (electron) out-of-plane flow $u_{iz}$ ($u_{ez}$), and out-of-plane current density $J_z$, through the X-point across the current sheet (CS) for both the initial and steady states.

According to Harris (1962), at the initial state, we set up the electron and ion drifting velocities at the center of CS to form the current, and the drifting velocities are proportional to the temperatures of corresponding species, $\frac{u_{i,CS}}{u_{e,CS}} = -\frac{T_i}{T_e}$. The out-of-plane bulk flow velocities are $u_{iz} = u_{i,msh} + u_{i,CS}$ and $u_{ez} = u_{e,msh} + u_{e,CS}$, where $u_{i,msh} = u_{e,msh} = u_{msh}$ are the out-of-plane magnetosheath flow, and $u_{i,CS} = \frac{T_i}{T_i+T_e}\frac{c}{4\pi e n}\nabla \times \boldsymbol{B}$ and $u_{e,CS} = -\frac{T_e}{T_i+T_e}\frac{c}{4\pi e n}\nabla \times \boldsymbol{B}$ are the drifting velocities. At t=0, $\nabla \times \boldsymbol{B} \approx -\partial B_x/\partial y$. As shown in Figure 6(a)-(e), for the dusk-side CS, $\nabla \times \boldsymbol{B} >$ 0, $u_{i,CS} > 0$, $u_{e,CS} < 0$; for the dawn-side CS, $\nabla \times \boldsymbol{B} < 0$, $u_{i,CS} < 0$, $u_{e,CS} > 0$. For the $M_A = 2.3$ (dusk) case, the ions are the major current carrier, while for other cases, both the ions and the electrons are significant for carrying the current. Note that the $x$-axis in Figure 6 is $y' = y - y_0$ for the dusk-side cases, and $y' = y_0 - y$ for the dawn-side cases. This arrangement is because all panels are organized with the magnetosheath on the left and the magnetosphere on the right. The upper and lower CSs have opposite orientations of the $y$-axis in the simulation domain, from magnetosheath to magnetosphere. The only caution is that $\frac{\partial}{\partial y} = \frac{\partial}{\partial y'}$ ($\frac{\partial}{\partial y} = -\frac{\partial}{\partial y'}$) for the dusk-side (dawn-side) cases. The same rule applies to Figures 7 and 8.

During the steady state, as shown in Figures 6(f)-(j), the electrons become the major current carrier in all cases. This is because the current sheet thickness is comparable to the scale of IDR, in which ions are decoupled from the magnetic field and they have lower mobility than electrons due to the mass. It is worth noting that the inflow ions and electrons from the magnetosheath side maintain the out-of-plane momentum (i.e., $u_{i,msh} = u_{e,msh} = u_{msh}$). When they reach the IDR, the electrons can easily change their momentum to form the current, while the ions experience a slow transition in their speed. For the $M_A = 2.3$ (dusk) case, the ion current is significant due to the same direction of the upstream out-of-plane ion bulk flow as the current direction. For the $M_A = 2.3$ (dawn) case, the direction of the ion upstream out-of-plane bulk flow is opposite to the current direction, which reduces the out-of-plane ion in both the upstream region and the current sheet. As shown in Figure 6(j), $u_{iz} \approx 1 = u_{msh}$ at $y' = -4.5$, while near the upstream edge of IDR, $u_{iz} \approx 0.61$ at $y' \approx -0.5$, which is sub-Alfvénic if normalized to the predicted outflow Alfvén speed $v_{out0} = 0.8$. As a result, this leads to an effectively slower upstream magnetosheath flow for the $M_A = 2.3$ (dawn) case. The reduction of $u_{iz}$ at the upstream edge is due to the electric force. Based on the ion momentum equation in a steady state, $nm_i u_{iy} \frac{\partial}{\partial y} u_{iz} \approx ne \left( E_z + \frac{u_{ix} B_y - u_{iy} B_x}{c} \right) = ne E_z'$, i.e., $\frac{\partial}{\partial y} u_{iz} \approx \frac{e E_z'}{m_i u_{iy}}$, where we ignore the pressure tensor term, and the $E_z'$ is the reconnection electric field in the ion moving frame which is equivalent to the reconnection rate. We consider $u_{iy} = \frac{J_y - J_{ey}}{en} \approx \frac{J_y}{en}$, where we ignore the electron current density because it is strong near the separatrices instead of the $x = x_0$ cut. Based on Ampere's law $J_y = -\frac{c}{4\pi} \frac{\partial}{\partial x} B_z$, and $\frac{\partial}{\partial x} B_z \approx \frac{B_{z,out} - (-B_{z,out})}{2L} = \frac{B_{z,out}}{L}$, we have $\frac{\partial}{\partial y} u_{iz} \approx -\frac{4\pi n e^2}{m_i} \frac{E_z'}{c} \frac{L}{B_{z,out}}$. When L is small and $B_{z,out}$ is large, $\frac{\partial}{\partial y} u_{iz}$ is close to zero and thus $u_{iz}$ is close to a constant, which is consistent with the $u_{iz}$ profile at the upstream edge ($y' \approx -0.5$) for the dusk-side cases (Figure 6(g) and (i)). When L is large and $B_{z,out}$ is small, $u_{iz}$ has a negative gradient toward the X-point. This explains the reduction of the upstream ion out-of-plane flow ($y' \approx -0.5$) for the dawn-side cases (Figure 6(h) and (j)). Note that before entering the IDR, the profile of electron out-of-plane flow $u_{ez}$ is the same as $u_{iz}$ since both the electrons and ions are frozen-in.

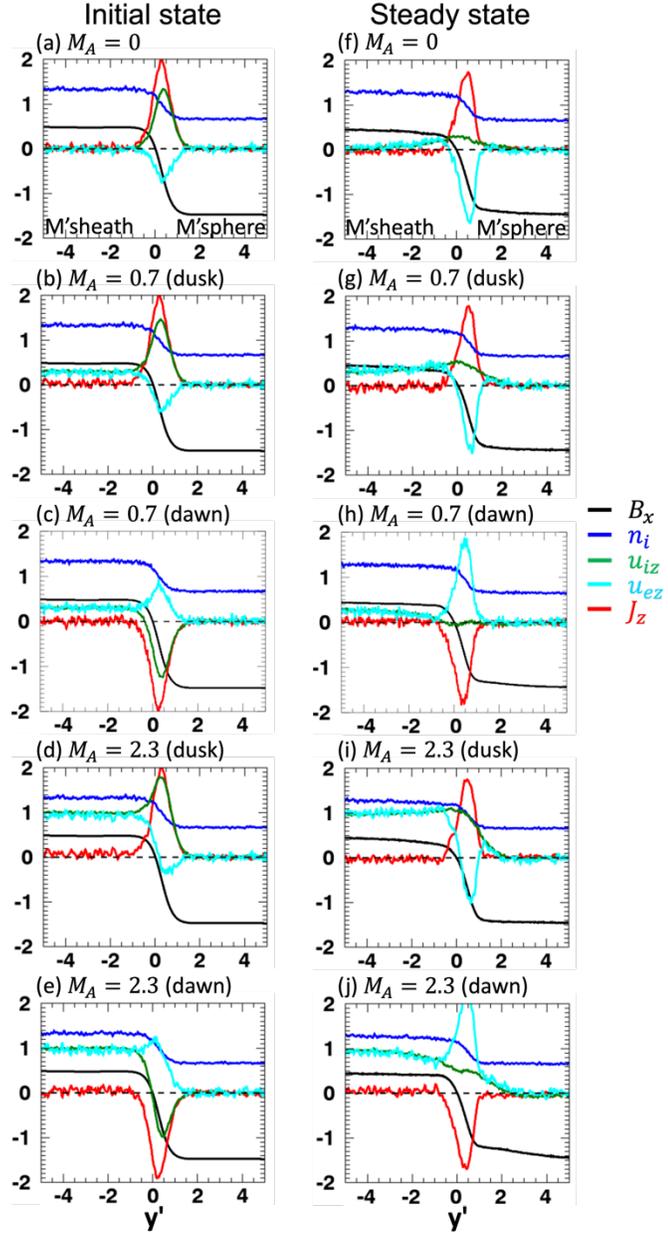

Figure 6 The profiles of $B_x$, $n_i$, $u_{iz}$, $u_{ez}$ and $J_z$ along y direction through the X-point during the initial and steady state for all the cases. The X-point is at $y = y_0$. The magnetosheath (magnetosphere) is on the left (right) of each panel. The horizontal axis is $y' = y - y_0$ for the dusk-side cases, and $y' = y_0 - y$ for the dawn-side cases.

3.4 Energy Conversion

We examine the impact of the out-of-plane flow shear on the energy conversion from the electromagnetic field to the plasma. Figure 7(a1)-(e1) show $\mathbf{J} \cdot \mathbf{E}'$ (e.g., Zenitani et al., 2011) and its components for all the cases. The peak values of $\mathbf{J} \cdot \mathbf{E}'$ are 0.11 for the $M_A = 2.3$ (dusk) case, 0.07 for the $M_A = 2.3$ (dawn) case and the $M_A = 0.7$ (dusk) case, and 0.05 for the $M_A = 0.7$ (dawn) case and the $M_A = 0$ case. The results indicate the out-of-plane flow shear can increase the energy conversion rate, although the energy conversion rate is lower at the dawn side than the dusk side. The figure also shows that when the out-of-plane flow shear is small, the contribution to $\mathbf{J} \cdot \mathbf{E}'$ is mainly $J_z E_z'$; when the out-of-plane flow shear is large, $J_y E_y'$ also has a significant contribution. For all cases, the major $J_z E_z'$ contribution is in $0 < y' < 1$. For the $M_A = 0.7$ (dusk) and $M_A = 2.3$ (dusk) cases, the $J_y E_y'$ has a positive contribution at $y' = 0$ and a negative contribution in $0 < y' < 1$.

To understand $J_y E_y'$, we analyze $J_y$ and $E_y'$. For $E_y'$, by examining the terms in the electron momentum equation, we found that $E_y' \approx E_y + \frac{u_{ez} B_x}{c} \approx -\frac{1}{n} \frac{\partial P_{e,yy}}{\partial y}$ and the other terms are ignorable. This is validated in Figure 7(a2)-(e2). The pressure term $-\frac{1}{n} \frac{\partial P_{e,yy}}{\partial y}$ shows two peaks with different signs at $y' = 0$ and $0 < y' < 1$. This is because $P_{e,yy}$ has a bump in $-0.3 < y' < 0.7$ as shown in Figure 8(c). The bump of $P_{e,yy}$ is due to the enhanced energy from the boost velocity caused by the Hall electric field $E_y$ (e.g., Bessho et al., 2016). We find that the profiles of $E_y + \frac{u_{ez} B_x}{c}$ and $-\frac{1}{n} \frac{\partial P_{e,yy}}{\partial y}$ show similar magnitudes for all cases because the local physics is the same. The quantity $J_y$ is thus the main factor that determines the difference of $J_y E_y'$ across all cases. As shown in Figure 7(a2)-(e2), $|J_y|$ is large near $y' = 0$ for the $M_A = 2.3$ (dusk) and $M_A = 0.7$ (dusk) cases, while it is small for the $M_A = 2.3$ (dawn) and $M_A = 0.7$ (dawn) cases. Because $|J_y| \propto \left| \frac{\partial B_z}{\partial x} \right| \sim \left| \frac{\Delta B_z}{2L} \right|$, the $B_z$ magnitude in the outflow region and the diffusion region length $L$ play important roles. As

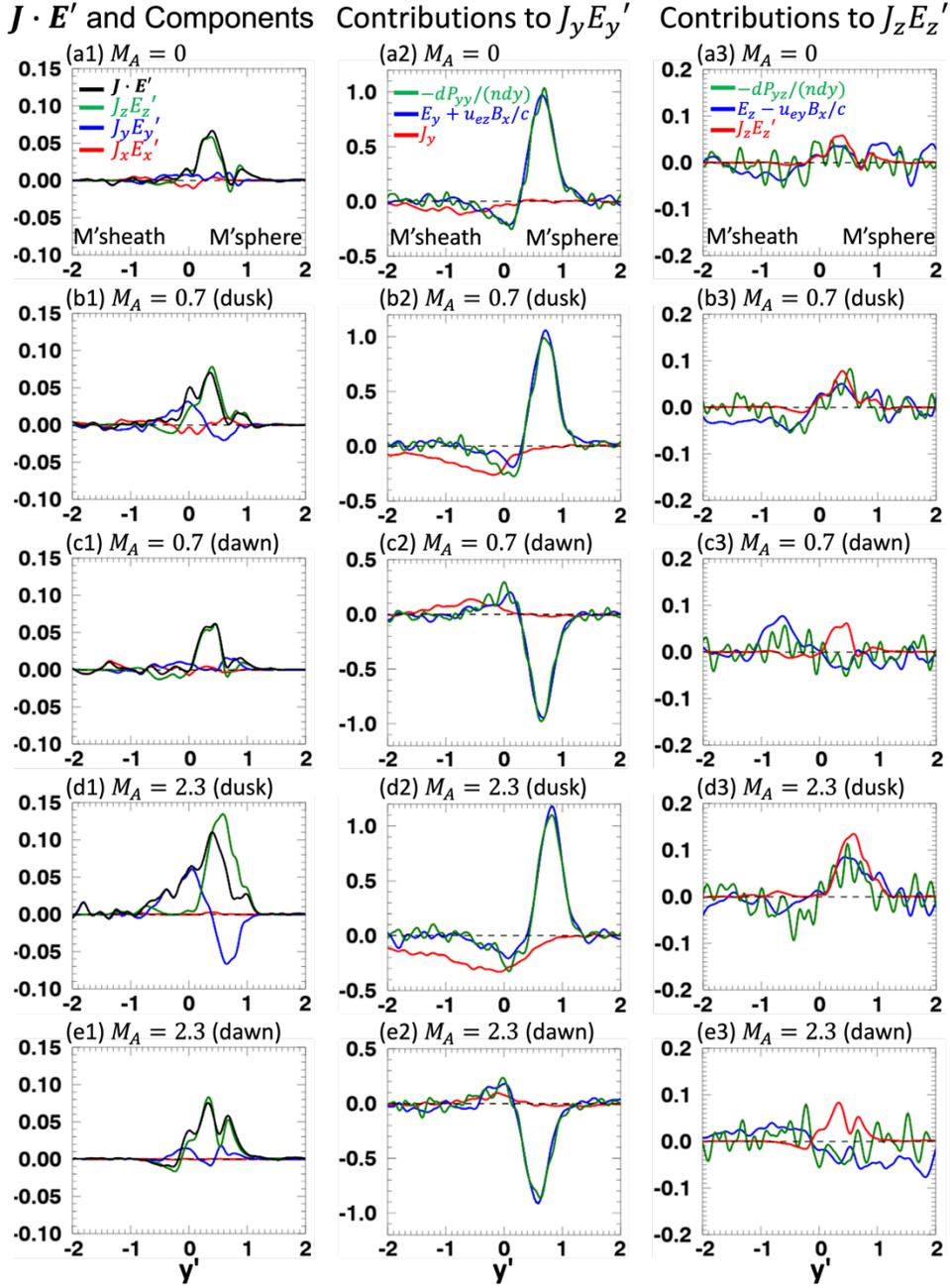

Figure 7 (a1)-(e1) $\mathbf{J} \cdot \mathbf{E}'$ and its components (a2)-(e2) Contributions to $J_y E_y'$ (a3)-(e3) Contributions to $J_z E_z'$. The cut is along the y direction through the X-point during the steady state for all the cases. The X-point is at $y = y_0$. The magnetosheath (magnetosphere) is on the left (right) of each panel. The horizontal axis is $y' = y - y_0$ for the dusk-side cases, and $y' = y_0 - y$ for the dawn-side cases.

discussed in Section 3.2, the out-of-plane flow drags the reconnected field lines along the z direction. For the dusk-side cases, the larger out-of-plane flow shear leads to the larger $B_z$ magnitude in the outflow region and the relevant magnetic pressure which squeezes the IDR along the outflow direction. In other words, for such cases, the quantity $|J_y|$ increases since $L$ decreases and $|\Delta B_z|$ increases, and this effect is proportional to the out-of-plane flow shear strength. For the dawn-side cases, the effect is opposite because as discussed in Section 3.2, the dragged reconnected field component cancels part of the Hall magnetic field, which reduces the $B_z$ as well as the magnetic pressure in the outflow region. Without squeezing by the magnetic pressure, the $L$ is large and the $|\Delta B_z|$ is small, which results in relatively small $|J_y|$. As a result, $J_y E_y'$ is significant for the dusk-side cases and is proportional to the out-of-plane flow shear strength, while it is smaller for the dawn-side cases.

For $J_z E_z'$, $E_z' \approx E_z - \frac{u_{ey} B_x}{c} \approx -\frac{1}{n}\frac{\partial P_{e,yz}}{\partial y}$. The profiles along $y$ through the X-point are shown in Figure 7(a3)-(e3). The value of $-\frac{1}{n}\frac{\partial P_{e,yz}}{\partial y}$ in $0 < y' < 1$ is largest for the $M_A = 2.3$ (dusk) case, and its value in other cases is similar. As shown in Figures 6(f)-(j), $|J_z|$ is similar for all cases, and the differences in the magnitude of $J_z E_z'$ in these cases are thus mainly due to $\left|\frac{1}{n}\frac{\partial P_{e,yz}}{\partial y}\right|$.

A question arises why $\left|\frac{1}{n}\frac{\partial P_{e,yz}}{\partial y}\right|$ is relatively large in the dusk side case when there is a magnetosheath flow. To address this question, we study the local 2D velocity distribution functions (VDFs) in the $v_y - v_z$ plane. In Figure 8(a)-(d), we compare $u_{ey}$, $u_{ez}$, $P_{e,yy}$, and $P_{e,yz}$ for all the cases. The $u_{ey}$ profiles indicate that the electrons carry most of the current $J_y$ in the ion diffusion region. The $u_{ez}$ shows the magnetosheath out-of-plane flow in $y' < -0.5$, and the out-of-plane flow that forms the current in $-0.5 < y' < 1$. Note that the $u_{ez}$ directions are opposite (same) between the magnetosheath flow and the current carrying flow for the dusk (dawn) cases. The $P_{e,yy}$ profiles are similar in all the cases, while $P_{e,yz}$ shows a larger gradient along $y$ for the dusk-side cases in $0 < y' < 0.5$ than the other cases. At locations $y' = -0.4$, $0.1$, and $0.5$, the 2D electron VDFs in the $v_y - v_z$ plane are calculated by integrating over $v_x$. For each VDF, the electrons are collected in a $\Delta x \times \Delta y = 1 \times 0.25$ rectangle region. At $y' = -0.4$, the VDFs for $M_A = 0$, $M_A = 2.3$ (dusk) and $M_A = 2.3$ (dawn) are shown in Figures 8(e1)-(e3). The distributions are similar for the three cases except for a drift of the electron bulk flow in the $v_y - v_z$ plane (blue line). In Figures 8(f1)-(f3), at $y' = 0.1$, the VDFs show a "U" shape embracing the origin in the perpendicular plane which is the meandering motion signature (e.g., Lapenta et al., 2016; Bessho et al., 2017). Note that the asymmetry with respect to $v_y = 0$ is due to the acceleration of the meandering electrons by the reconnection electric field $E_z$ (Bessho et al., 2017). This asymmetry results in the non-zero $P_{e,yz}$. Because $P_{e,yz} \propto \int (v_y - u_{ey})(v_z - u_{ez}) f(v_y, v_z) dv_y dv_z = \int v_y' v_z' f(v_y', v_z') dv_y' dv_z'$, the value of $P_{e,yz}$ is determined by the positive contribution where $v_y' v_z' > 0$ and the negative contribution where $v_y' v_z' < 0$, where $v_y' = v_y - u_{ey}$ and $v_z' = v_z - u_{ez}$. At $y' = 0.1$, $u_{ey} \approx 0$ and $u_{ez} \approx J/n$, representing the current-carrying electron flow. For the "U" shape distribution shown in Figure 8(h1)-(h2) for the $M_A = 0$ and $M_A = 2.3$ (dusk) cases, the integration of $v_y' v_z' f(v_y', v_z')$ over $(v_y' < 0, v_z' < 0)$ is approximately canceled by the integration over $(v_y' > 0, v_z' < 0)$ due to the opposite sign. Therefore, $P_{e,yz}$ is mainly determined by the region $(v_y' > 0, v_z' > 0)$ since the phase space density in $(v_y' < 0, v_z' > 0)$ is relatively small. According to Bessho et al. (2017), the region $(v_y' > 0, v_z' > 0)$ contains particles starting meandering motion with upstream velocity $v_0$. Given that the local electric field is similar in the compared cases, based on the Eqn. (33) in Bessho et al. (2017), the distance of the particle trajectory in ($v_y' > 0, v_z' > 0$) to the origin is approximately proportional to the initial speed of the meandering motion $v_0$. The larger distance of the trajectory in ($v_y' > 0, v_z' > 0$) to the origin results in the larger integral of $v_y' v_z' f(v_y', v_z')$.

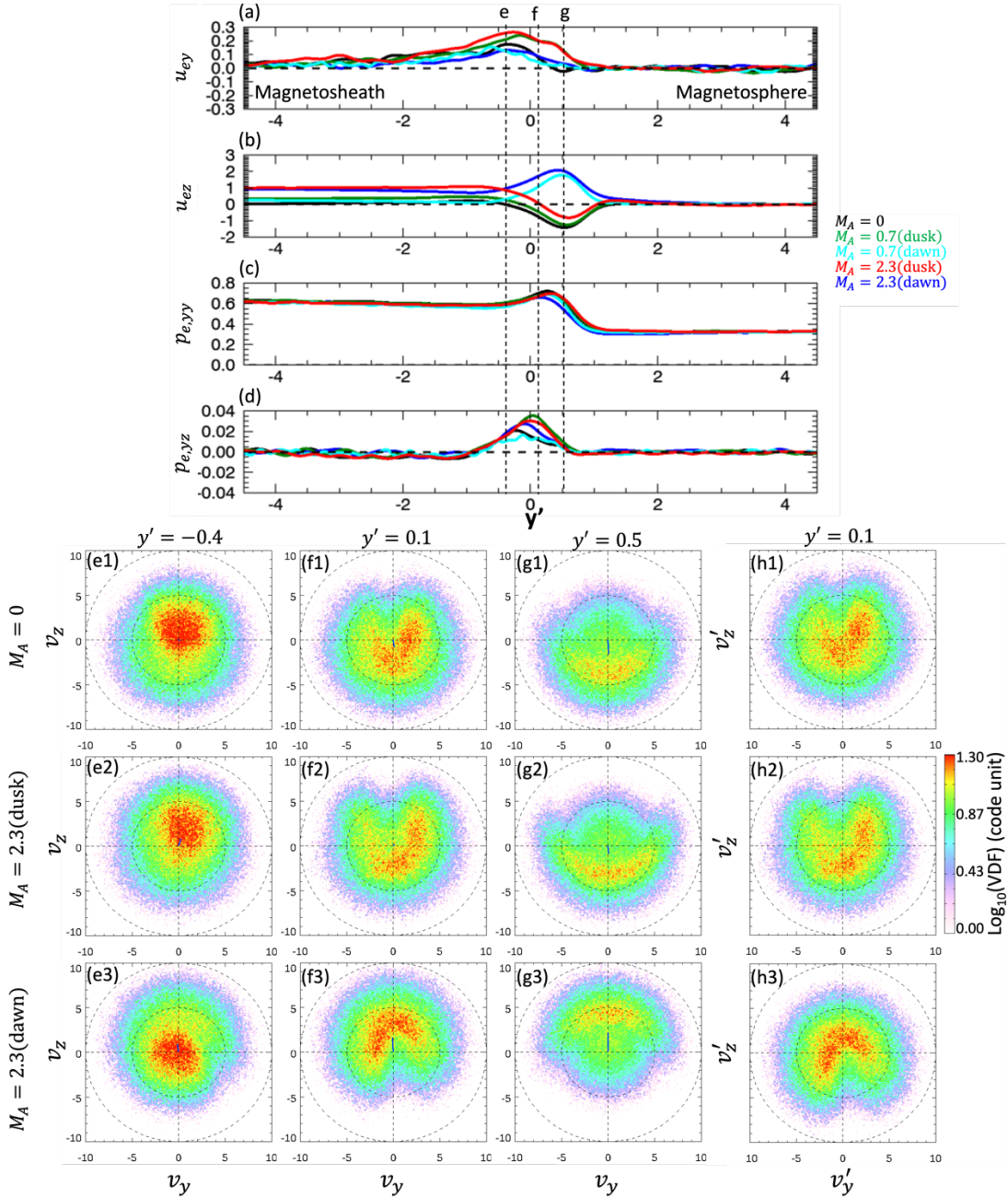

Figure 8 (a)-(d) The profiles of $u_{ey}$, $u_{ez}$, $P_{e,yy}$, and $P_{e,yz}$ along y through the X-point during the steady state for all the cases. The X-point is at $y = y_0$. The magnetosheath (magnetosphere) is on the left (right) of Panels (a)-(d). The horizontal axis is $y' = y - y_0$ for the dusk-side cases, and $y' = y_0 - y$ for the dawn-side cases. The vertical dashed lines indicate the locations ($y' = -0.4$, 0.1, and 0.5) of the 2D VDFs shown in (e1)-(e3), (f1)-(f3), and (g1)-(g3) for the $M_A = 0$, $M_A = 2.3$ (dusk) and $M_A = 2.3$ (dawn) cases. The VDFs are in the $v_y - v_z$ plane after integration over $v_x$. The blue line in the 2D VDF plots indicates the bulk flow velocity ($u_{ey}, u_{ez}$). (h1)-(h3) 2D VDF at $y' = 0.1$ in bulk flow velocity frame, where $v_y' = v_y - u_{ey}$ and $v_z' = v_z - u_{ez}$.

Since the $v_0$ is larger in the $M_A = 2.3$ (dusk) case due to the additional upstream out-of-plane flow component, the $P_{e,yz}$ in this case is larger compared to the $M_A = 0$ case. A similar analysis is applied to the Figure 8(h3), in which the $P_{e,yz}$ is mainly determined by the integration over ($v_y' < 0, v_z' < 0$). Although the $M_A = 2.3$ (dawn) case includes an upstream out-of-plane flow contribution to $v_0$ as well, the upstream out-of-plane flow is reduced as shown in Figure 6 and discussed in the Section 3.3. Therefore, the $P_{e,yz}$ value for the $M_A = 2.3$ (dawn) case is between the $M_A = 2.3$ (dusk) case and the $M_A = 0$ case. At $y' = 0.5$, all the three cases show symmetric crescent shape distributions (e.g., Shay et al., 2016; Bessho et al., 2016; Lapenta et al., 2016; Egedal et al., 2016; Hesse et al., 2021) and thus $P_{e,yz} \approx 0$. As a result, by examining the $P_{e,yz}$ gradient between $y' = 0.1$ and 0.5, we found that the magnetosheath flow in the dusk-side reconnection results in larger $\left|\frac{\partial P_{e,yz}}{\partial y}\right|$ than that in the dawn-side reconnection.

## 4. Discussion and Conclusions

In this work, we have studied the effect of the out-of-plane flow shear on asymmetric reconnection occurring at the flanks of Earth's magnetopause. Here, we have assumed the flow in the magnetosphere is close to zero and the out-of-plane flow shear is produced by the magnetosheath flow along the solar wind direction. Under the southward IMF condition, the reconnection current sheet direction is same as (opposite to) the magnetosheath flow on the dusk (dawn) side. We have used 2.5D PIC simulations of dusk- and dawn- side reconnection to study the reconnection rate, the diffusion region configuration, and the energy conversion under the magnetosheath flows with the Alfvén Mach numbers $M_A = 0, 0.7$, and 2.3, which correspond to no flow, sub-Alfvénic, and super-Alfvénic flows, respectively. We have found that:

(1) The reconnection rate increases with the upstream out-of-plane flow shear, and for the same flow shear, it is higher at the dusk side than at the dawn side. The reconnection rate is qualitatively consistent with the aspect ratio $\delta/L$ of the IDR.

(2) The out-of-plane flow drags the reconnected field lines, adding the out-of-plane magnetic field component, which provides magnetic pressure ($\propto (M_A^{CS07})^2$) to compress the IDR length $L$, where $M_A^{CS07}$ is the Alfvén Mach number for the out-of-plane magnetosheath flow $u_{msh}$ normalized to the predicted outflow Alfvén speed from Cassak and Shay (2007).

(3) The out-of-plane flow shear increases the energy conversion rate $\boldsymbol{J} \cdot \boldsymbol{E}'$. For the same upstream out-of-plane flow shear, the energy conversion rate $\boldsymbol{J} \cdot \boldsymbol{E}'$ is higher at the dusk side than the dawn side.

Our simulations with super-Alfvénic flows do not show a reversed current at the reconnection site as Liu et al. (2018). One reason is that the super-Alfvénic out-of-plane flow in our simulation is not sufficiently strong to provide the necessary magnetic pressure in the outflow region as in Liu et al. (2018). It is worth noting that in Liu et al. (2018), it is theoretically possible that the super-Alfvénic condition for reversing the X-line current in 'pair plasmas' could be interpreted as the super-electron-Alfvénic condition in electron-proton plasmas. Thus, this interpretation might suggest that a higher flow shear is necessary to achieve the reversed X-line current in this study. Further investigation is required to establish a physical basis for this hypothesis. Another reason could be due to the asymmetric nature of magnetopause reconnection configuration. In Liu et al. (2018), the reversed current contribution is from the gradient of the reconnected magnetic field (the component normal to the current sheet) along the outflow direction, i.e., $\Delta B_y/2L$ using the coordinates and symbols in this paper. In asymmetric reconnection, the original current $\Delta B_x/2\delta$ is near the stagnation point close to the magnetosphere side, while the current due to $\Delta B_y/2L$ is mainly in the magnetosheath side since the outflow region expands toward the weak-field side. As a result, even though the reversed current can potentially occur under the super-Alfvénic out-of-plane flow, it is not able to reverse the original current direction unless the magnetic field asymmetry is weak.

In our study, we derived a relation $B_{z,out} \sim \frac{u_{msh}}{v_{out0}} \frac{b_1 b_2}{b_1 + b_2}$, which aligns closely with the findings of Sun et al. (2005). According to the reconnection layer theory (Lin and Lee, 1993) discussed by Sun et al. (2005), the formation of $B_{z,out}$ can be interpreted by replacing the pair of switch-off shocks in the Petschek reconnection model with a pair of slow shocks and a pair of time-dependent intermediate shocks (TDIS) or rotational discontinuities (RDs). These TDIS/RDs generate $B_{z,out}$ by following the Walen (1944) relation, $\Delta u_s = \pm \Delta v_A$, in the local de Hoffmann and Teller (1950) frame, where $\Delta u_s$ and $\Delta v_A$ are the changes in bulk velocity and Alfvén speed, respectively. Subsequently, $B_{z,out}$ undergoes slight modification by the pair of slow shocks. For a symmetric case, the de Hoffmann and Teller frame has no $u_z$ component. However, for an asymmetric case, the frame can move along the z direction, leading to $B_{z,out} \sim \Delta u_z$. Thus, using a different method, our result is consistent with the findings of Sun et al. (2005).

One of the major effects of the out-of-plane flow shear is that it can drag the reconnected field lines and increase (decrease) the Hall magnetic field in the outflow region for the dusk(dawn)-side case as discussed in Section 3.2. The increment (reduction) Hall magnetic field is proportional to the out-of-plane flow speed. One interesting question is whether an out-of-plane flow can be sufficiently strong and cancel the Hall magnetic field for the dawn-side reconnection. This question is important because the Hall magnetic field is associated with the Hall currents which are carried by the electrons that flow toward the X-point along the separatrices and then are ejected as outflow jets. On one hand, if the Hall magnetic field is canceled out, then the Hall-current-carrier electrons may stop flowing toward the X-point, which may suppress the reconnection. On the other hand, if the reconnection is suppressed, the impact of the out-of-plane flow on the Hall magnetic field would disappear since there are no more reconnected field lines to be dragged out. As a result, reconnection may reach a state with a sufficiently small reconnection rate that can balance the production of reconnected field lines and the reduction of the Hall magnetic field. Investigating the effect of faster out-of-plane flow, which may lead to a strong dragged component that could cancel the Hall magnetic field, is an interesting open question.

In this study, we consider a finite magnetosheath flow together with zero magnetospheric flow. The conclusions in this paper are still valid for a non-zero magnetospheric out-of-plane flow, because one can always transfer to a moving frame of the magnetospheric flow. Only the flow shear value matters to the conclusions in this work. In addition, we did not consider a guide field which could be quite common at the flanks. A guide field can twist the diffusion region configuration and change the current pattern near the reconnection site. Therefore, the conclusions may not work for the case with a guide field. Understanding the out-of-plane flow effect on the asymmetric reconnection with a guide field remains an interesting topic for further investigation. Furthermore, it is important to note that the conclusions of this study are based on 2D geometry. In a 3D context, out-of-plane flow shear can trigger the Kelvin-Helmholtz instability, potentially leading to significant modifications of the X-line configuration (Ma et al., 2014a, 2014b).

## Data Availability Statement

The PIC simulation code (P3D) used in this study is available online (https://terpconnect.umd.edu/~swisdak/p3d/). The output data of the simulation runs used in this production of all figures is available online for download at https://doi.org/10.5281/zenodo.13820419 (Liang et al., 2024).

## Acknowledgments

This study is supported by the NASA MMS Mission. HL acknowledges partial support from NASA grant 80NSSC24K0388 and NSF grant AGS2247718. This research used resources of the National Energy Research Scientific Computing Center (NERSC), a DOE Office of Science User Facility supported by the Office of Science of the U.S. Department of Energy under Contract No. DE-AC02-05CH11231, using NERSC award FES-ERCAP0028376. Additional resources were provided by NASA HECC Pleiades and NSF TACC Frontera.

## References

Bessho, N., Chen, L. J., & Hesse, M. (2016). Electron distribution functions in the diffusion region of asymmetric magnetic reconnection. Geophysical Research Letters, 43(5), 1828-1836.

Bessho, N., Chen, L. J., Hesse, M., & Wang, S. (2017). The effect of reconnection electric field on crescent and U-shaped distribution functions in asymmetric reconnection with no guide field. Physics of Plasmas, 24(7).

Birdsall, C. K., & Langdon, A. B. (2018). *Plasma physics via computer simulation*. CRC press.

Cassak, P. A., & Shay, M. A. (2007). Scaling of asymmetric magnetic reconnection: General theory and collisional simulations. Physics of Plasmas, 14(10).

Chen, Y., et al., (2013). The influence of out-of-plane shear flow on Hall magnetic reconnection and FTE generation. Journal of Geophysical Research: Space Physics, 118(7), 4279-4288.

Cowley, S. W. H., & Lockwood, M. (1992). Excitation and decay of solar wind-driven flows in the magnetosphere-ionosphere system. In Annales geophysicae (Vol. 10, No. 1-2, pp. 103-115).

De Hoffmann, F., & Teller, E. (1950). Magneto-hydrodynamic shocks. *Physical Review*, *80*(4), 692.

Egedal, J., Le, A., Daughton, W., Wetherton, B., Cassak, P. A., Chen, L. J., ... & Avanov, L. A. (2016). Spacecraft observations and analytic theory of crescent-shaped electron distributions in asymmetric magnetic reconnection. Physical review letters, 117(18), 185101.

Fuselier, S. A., et al. (2005), Computing the reconnection rate at the Earth's magnetopause using two spacecraft observations, J. Geophys. Res., 110, A06212, doi:10.1029/2004JA010805.

Goldman, M. V., Newman, D. L., & Lapenta, G. (2016). What can we learn about magnetotail reconnection from 2D PIC Harris-sheet simulations?. *Space Science Reviews*, *199*(1), 651-688.
Schindler, K., Hesse, M., & Birn, J. (1991). Magnetic field-aligned electric potentials in nonideal plasma flows. *Astrophysical Journal, Part 1 (ISSN 0004-637X), vol. 380, Oct. 10, 1991, p. 293-301. Research supported by DFG, DOE, and NASA.*, *380*, 293-301.

Gurram, H., Shuster, J. R., Chen, L. J., Argall, M. R., Denton, R. E., Rice, R. C., ... & Gershman, D. J. (2024). Turbulence properties and kinetic signatures of electron in Kelvin-Helmholtz waves during a geomagnetic storm. arXiv preprint arXiv:2403.18990.

Guzdar, P. N., Drake, J. F., McCarthy, D., Hassam, A. B., & Liu, C. S. (1993). Three-dimensional fluid simulations of the nonlinear drift-resistive ballooning modes in tokamak edge plasmas. *Physics of Fluids B: Plasma Physics*, *5*(10), 3712-3727.

Harris, E. G. (1962). On a plasma sheath separating regions of oppositely directed magnetic field. Il Nuovo Cimento (1955-1965), 23, 115-121.

Hasegawa, H., et al. (2016), Decay of mesoscale flux transfer events during quasi-continuous spatially-extended reconnection at the magnetopause, Geophys. Res. Lett., 43, 4755–4762, doi:10.1002/2016GL069225.

Hesse, M., Norgren, C., Tenfjord, P., Burch, J. L., Liu, Y.-H., Bessho, N., et al. (2021). A new look at the electron diffusion region in asymmetric magnetic reconnection. Journal of Geophysical Research: Space Physics, 126, e2020JA028456. https://doi.org/10.1029/2020JA028456

La Belle-Hamer, A. L., Otto, A., & Lee, L. C. (1995). Magnetic reconnection in the presence of sheared flow and density asymmetry: Applications to the Earth's magnetopause. *Journal of Geophysical Research: Space Physics*, *100*(A7), 11875-11889.

Lapenta, G., Berchem, J., Zhou, M., Walker, R. J., El-Alaoui, M., Goldstein, M. L., ... & Burch, J. L. (2017). On the origin of the crescent-shaped distributions observed by MMS at the magnetopause. *Journal of Geophysical Research: Space Physics*, *122*(2), 2024-2039.

Liang, H., Chen, L.-J., Bessho, N., & Ng, J. (2024). Data for "Impact of the Out-of-Plane Flow Shear on Magnetic Reconnection at the Flanks of Earth's Magnetopause" [Data set]. Zenodo. https://doi.org/10.5281/zenodo.13820419

Lin, Y., & Lee, L. C. (1993). Structure of reconnection layers in the magnetosphere. *Space Science Reviews*, *65*(1), 59-179.

Liu, Y. H., et al., (2018). Strongly localized magnetic reconnection by the super-Alfvénic shear flow. Physics of plasmas, 25(8).

Ma, X., Otto, A., & Delamere, P. A. (2014a). Interaction of magnetic reconnection and Kelvin-Helmholtz modes for large magnetic shear: 2. Reconnection trigger. *Journal of Geophysical Research: Space Physics*, *119*(2), 808-820.

Ma, X., Otto, A., & Delamere, P. A. (2014b). Interaction of magnetic reconnection and Kelvin-Helmholtz modes for large magnetic shear: 1. Kelvin-Helmholtz trigger. *Journal of Geophysical Research: Space Physics*, *119*(2), 781-797.

Ma, X., Otto, A., & Delamere, P. A. (2016). Magnetic reconnection with a fast perpendicular sheared flow. *Journal of geophysical research: Space physics*, *121*(10), 9427-9442.

Nakamura, T. K. M., et al. (2020). Decay of Kelvin-Helmholtz vortices at the Earth's magnetopause under pure southward IMF conditions. Geophysical Research Letters, 47(13), e2020GL087574.

Phan, T.-D., G. Paschmann, and B. U. Ö. Sonnerup (1996), Low latitude dayside magnetopause and boundary layer for high magnetic shear: 2. Occurrence of magnetic reconnection, J. Geophys. Res., 101, 7817.

Phan, T.-D., et al. (2000), Extended magnetic reconnection at the Earth's magnetopause from detection of bi-directional jets, Nature, 404, 848.

Pritchett, P. L. (2008). Collisionless magnetic reconnection in an asymmetric current sheet. *Journal of Geophysical Research: Space Physics*, *113*(A6).

Reistad, J. P., et al. (2018). Observations of asymmetries in ionospheric return flow during different levels of geomagnetic activity. Journal of Geophysical Research: Space Physics, 123(6), 4638-4651

Scurry, L., C. T. Russell, and J. T. Gosling (1994), A statistical study of accelerated flow events at the dayside magnetopause, J. Geophys. Res., 99, 14,815.

Shay, M. A., Phan, T. D., Haggerty, C. C., Fujimoto, M., Drake, J. F., Malakit, K., ... & Swisdak, M. (2016). Kinetic signatures of the region surrounding the X line in asymmetric (magnetopause) reconnection. Geophysical Research Letters, 43(9), 4145-4154.

Souza, V. M., Gonzalez, W. D., Sibeck, D. G., Koga, D., Walsh, B. M., & Mendes, O. (2017). Comparative study of three reconnection X line models at the Earth's dayside magnetopause using in situ observations. Journal of Geophysical Research: Space Physics, 122(4), 4228-4250.

Sun, X., Lin, Y., & Wang, X. (2005). Structure of reconnection layer with a shear flow perpendicular to the antiparallel magnetic field component. Physics of plasmas, 12(1).

Swisdak, M. (2016). Quantifying gyrotropy in magnetic reconnection. Geophysical Research Letters, 43(1), 43-49.

Trattner, K. J., J. S. Mulcock, S. M. Petrinec, and S. A. Fuselier (2007), Probing the boundary between antiparallel and component reconnection during southward interplanetary magnetic field conditions, J. Geophys. Res., 112, A08210, doi:10.1029/2007JA012270.

Trottenberg, U., & Clees, T. (2009). Multigrid software for industrial applications-from MG00 to SAMG. In *100 Volumes of 'Notes on Numerical Fluid Mechanics' 40 Years of Numerical Fluid Mechanics and Aerodynamics in Retrospect* (pp. 423-436). Berlin, Heidelberg: Springer Berlin Heidelberg.

Walén, C. (1944). *On the theory of sun-spots* (Doctoral dissertation, Almqvist & Wiksell).

Wang, J., et al. (2008). Out-of-plane bipolar and quadrupolar magnetic fields generated by shear flows in two-dimensional resistive reconnection. Physics Letters A, 372(25), 4614-4617.

Wang, J., et al. (2012). Effects of out-of-plane shear flows on fast reconnection in a two-dimensional hall magnetohydrodynamics model. Physics of Plasmas, 19(3).

Wang, L., et al. (2015). Asymmetric magnetic reconnection with out-of-plane shear flows in a two dimensional hybrid model. Physics of Plasmas, 22(5).

Zeiler, A., et al. (2002). Three-dimensional particle simulations of collisionless magnetic reconnection. Journal of Geophysical Research: Space Physics, 107(A9), SMP-6

Zenitani, S., Hesse, M., Klimas, A., & Kuznetsova, M. (2011). New measure of the dissipation region in collisionless magnetic reconnection. *Physical review letters*, *106*(19), 195003.

Zou, Y., Walsh, B. M., Nishimura, Y., Angelopoulos, V., Ruohoniemi, J. M., McWilliams, K. A., & Nishitani, N. (2018). Spreading speed of magnetopause reconnection X‐lines using ground‐satellite coordination. Geophysical Research Letters, 45(1), 80-89.